\documentclass[journal=jacsat,manuscript=article]{achemso}

\usepackage{amsmath}
\usepackage{amsmath,amssymb}
\usepackage{graphicx}
\usepackage{caption}
\usepackage{color}
\usepackage{dcolumn}
\usepackage{bm}
\usepackage{float}
\usepackage{subfig}
\usepackage{soul}
\usepackage{comment}
\usepackage[normalem]{ulem}
\usepackage{microtype}

\usepackage{rotating}

\SectionNumbersOn

\author{Eric Boittier} \affiliation[University of Basel]{Department
of Chemistry, University of Basel, Klingelbergstrasse 80 , CH-4056
Basel, Switzerland.}

\author{Kai T\"opfer} \affiliation[University of Basel]{Department
of Chemistry, University of Basel, Klingelbergstrasse 80 , CH-4056
Basel, Switzerland.}

\author{Mike Devereux} \affiliation[University of Basel]{Department
of Chemistry, University of Basel, Klingelbergstrasse 80 , CH-4056
Basel, Switzerland.}

\author{Markus Meuwly} \affiliation[University of Basel]{Department of
  Chemistry, University of Basel, Klingelbergstrasse 80 , CH-4056
  Basel, Switzerland.}  \email{m.meuwly@unibas.ch}

\title{Kernel-based Minimal Distributed Charges: A Conformationally
  Dependent ESP-Model for Molecular Simulations}

\date{\today}
\begin{document}

\begin{abstract}
A kernel-based method (kernelized minimal distributed charge model -
kMDCM) to represent the molecular electrostatic potential (ESP) in
terms of off-center point charges whose positions adapts to the
molecular geometry. Using Gaussian kernels and atom-atom distances as
the features, the ESP for water and methanol is shown to improve by at
least a factor of two compared with point charge models fit to an
ensemble of structures.  Combining kMDCM for the electrostatics and
reproducing kernels for the bonded terms allows energy-conserving
simulation of 2000 water molecules with periodic boundary conditions
on the nanosecond time scale.
\end{abstract}

\section{Introduction}
Empirical energy functions (EEFs) constitute an important framework in
state-of-the art characterization of the energetics and dynamics of
molecular systems in the gas- and the
condensed-phase.\cite{shim:2011,gunsteren:2018,MM.rev:2022} The
success of EEFs is primarily owed to the speed with which energies and
forces can be evaluated for a given arrangement of the atoms. In the
most conventional formulation, EEFs distinguish between bonded and
nonbonded energy contributions and the nonbonded interactions
encompass van der Waals and electrostatic interactions.\cite{cgenff}
The electrostatic contributions are often conveniently described by
atom-centered point charge (PC) models which can be considered as the
first term of a multipolar expansion based on atom-centered
moments.\cite{stone:2013} \\

\noindent
It has been long recognized that correctly capturing the anisotropy of
the electrostatic (Coulomb) contributions to the total energy is
mandatory for a realistic description of the intermolecular
interactions.\cite{rein:1973} One possibility to achieve this is to
represent the molecular electrostatic potential (ESP) as a
superposition of atom-centered multipoles (MTPs) to given
order. Often, this expansion is truncated at the atomic quadrupole
moment for all atoms except hydrogens for which a PC representation is
usually
sufficient.\cite{Cisneros:2006,MM.mbco:2008,ponder:2010,Kramer:2014}
Alternatively, PC-based energy functions can be supplemented by
off-center PCs to model the anisotropy of the ESP originating from an
anisotropic electron density due to, for example,
$\sigma-$holes\cite{schyman:2012} or lone pairs.\\

\noindent
Besides charge anisotropy, intramolecular charge redistribution occurs
as a consequence of changes in molecular geometry.\cite{soos:2014}
Such changes in the electron density lead to fluctuations in the
electric dipole and higher molecular moments which affect, among
others, the intensities of infrared (IR) spectra. Modelling such
effects can be accomplished through conformationally dependent
atom-centered multipoles.\cite{MM.mbco:2008,stone:1995,price:1990} It
was also found that the electric field around a molecule depends on
all molecular degrees of freedom,\cite{Jensen:2022, Sedghamiz:2017}
which cannot be described by simply translating and rotating a locally
frozen electron density to a new spatial position and orientation. As
an example, the molecular dipole moment of water is often cited to
demonstrate the conformational dependence of the electrostatic
potential,\cite{Dinur:1990, Marechal:2011, Xantheas:2006} as
fluctuations in the range of $\sim 50\%$ have been observed between
isolated molecules and the condensed and solid
phases.\cite{Ren:2020}\\

One solution to include conformational dependence into charge models
has been to scale atom-centered charges to reproduce the
conformationally averaged molecular dipole moment. However, such a
``mean field'' ansatz has no physical basis and can lead to artifacts
in the arrangement of molecules within solvation
shells.\cite{Bedrov:2019} Several fluctuating atomic PC models, which
scale the magnitude of the charges by a function depending on
conformation have been developed.\cite{Kim:2019,Reynolds:1992}
However, they are limited in describing charge anisotropy without
higher-order multipoles\cite{Richter:2022}.\\

\noindent
Another class of off-center PC representations that has been developed
over the past decade which include the distributed charge model
(DCM)\cite{MM.dcm:2014}, minimal distributed charge model
(MDCM)\cite{MM.mdcm:2017,MM.mdcm:2020} or
others.\cite{Liu:2020,Cools:2021,Paesani:2023} Off-center models
require local reference frames (sets of three non-collinear atoms) to
define the charge positions given rotations and translations of the
molecule.\cite{MM.mtp:2012,MM.dcm:2014} The choice of local axis
systems is non-unique and can have subtle implications on the
performance of the model.\cite{Liedl:2024} Fluctuating minimal
distributed charge models (fMDCM) allow to describe changes in the
molecular ESP depending on molecular structure.\cite{MM.fmdcm:2022}
This represents intramolecular charge redistribution (polarization) as
a consequence of geometry changes. Finally, external polarization has
been investigated by ``charge-on-a-spring''
models.\cite{straatsma:1990,Yu:2005,lopes:2013} The Drude model
represents electronic induction by using the displacement of a
charge-carrying massless particle attached to an atom which changes
positions harmonically in response to the local electric field
originating from the surrounding charge distribution. \\

\noindent
Although fMDCM rather successfully described changes in the ESP
depending on conformation, generalizing such an approach to include
more or all internal degrees of freedom is not straightforward. In
particular, choosing and fitting a suitable parametrized function to
cast the geometry dependence of the MDCM charges can be challenging
for larger molecules.\cite{MM.fmdcm:2022} In these regards, machine
learning-based models have opened new avenues because they can be
understood as general function approximators. Two particularly
successful formulations are based on neural networks (NNs) and
kernels.\cite{Cools:2021,gp06} Kernel-based methods have long been
recognized to be ideally suited to address challenging fitting
problems, for example for intermolecular potential energy
surfaces.\cite{rabitz:1996,MM.rkhs:2017,sauceda:2019} Thus, a viable
approach is to explore kernel-based representations of the
off-centered charge positions within MDCM to describe their geometry
dependency of the off-centered charge positions in order to capture
changes in the ESP and the molecular dipole moment.\\

\noindent
The present work introduces kernel-based minimal distributed charge
model (kMDCM) for describing intramolecular charge redistribution
depending on molecular conformation. The formalism is applied to water
and methanol and energy-conserving trajectories on the nanosecond time
scale for a box of water molecules is carried out. The manuscript is
structured as follows. First, the formal and computational methods are
presented. This is followed by validation of kMDCM, comparison with
alternative and related methods and applications to condensed-phase
simulations.

\section{Methods}

\subsection{Kernel-Based Minimal Distributed Charges}
Kernel-based MDCM (kMDCM) optimally parametrizes the positions
$\boldsymbol{\delta}$ of $N$ charges depending on molecular
geometry. Here, $N$ is the number of MDCM charges chosen for
describing the molecular ESP. The initial charge positions
$\boldsymbol{\delta}_0$ are those from a conformationally averaged
MDCM model.\cite{MM.mdcm:2017} The loss function
\begin{equation}
    \mathcal{L}(\boldsymbol{\delta}) =
    \mathrm{RMSE}(\boldsymbol{\delta}) + \lambda_1 \cdot
    ||\boldsymbol{\delta} -\boldsymbol{\delta_0}||^{2}_{2}
    \label{eq:lrmse}
\end{equation}
reduces the root-mean-squared-error
$\mathrm{RMSE}(\boldsymbol{\delta})$ between the reference molecular
ESP$_{\rm ref.}$ obtained from quantum chemical calculations and the
ESP from the MDCM charges positioned at $\boldsymbol{\delta}$ for
molecular geometry $\mathbf{R}$, and evaluated over $N_{\rm grid}$
grid points $\mathbf{y}_i$
  \begin{equation}
      \mathrm{RMSE}(\boldsymbol{\delta}) :=
      \sqrt{
        \frac{1}{N_{\rm grid}}
        \sum^{N_{\rm grid}}_{i=1}
          \left [ 
            {\rm ESP}_{\rm ref.}(\mathbf{y}_i, \mathbf{R}) 
            - {\rm ESP}_{\rm MDCM}(\mathbf{y}_i; \boldsymbol{\delta}, \mathbf{R})
          \right ] ^{2}}
  \end{equation}
Here, the MDCM ESP is calculated as the sum of the Coulomb
interactions between each kMDCM charge and a probe charge of 1$e$
placed on each of the $N_{\rm grid}$ grid points. For charge positions
$\boldsymbol{\delta}$ in the global reference system,
$\mathcal{L}(\boldsymbol{\delta})$ was optimized using
l-BFGS.\cite{Liu:1989} To afford greater control on the displacements
of the off-center atom charges, a penalty term scaled by a hyperparameter,
$\lambda_1$ was added to the loss function which penalises the
solution relative to the initial value positions
$\boldsymbol{\delta_0}$ in Eq. \ref{eq:lrmse}.\\

Within kMDCM, the MDCM charge positions for $N_{\rm train}$ training
structures are represented as a kernel matrix $K$ based on Gaussian
kernel functions. Given two structures $\mathbf{R}$ and $\mathbf{R}'$
for a molecule with $N_{\rm a}$ atoms, the input (or features) for
evaluating the elements of the kernel matrix $K(\mathbf{d,d'})$ are
the $N_{\rm a}\times(N_{\rm a}-1)/2-$dimensional interatomic distance
vectors $\mathbf{d}$ and $\mathbf{d}'$ to evaluate
\begin{equation}
      K(\mathbf{d,d'}) = \exp\Bigg(-\frac{\sqrt{\sum_{i=2}^{N_{\rm a}}\sum_{j<i}^{N_{\rm a}}(d_{ij} -
          d'_{ij})^{2}}}{2\sigma^{2}}\Bigg)
    \label{eq:gauskern}
\end{equation}
where $d_{ij}$ is the distance between atoms $i$ and $j$, and
$\sigma=1.0$ \AA\/ is the scale length of the kernel. For a triatomic
molecule, the list of internal distances $\mathbf{d}$ contains
$d_{12}, d_{13}, d_{23}$ where, for example, $d_{12}$ is the distance
between atoms 1 and 2.\\

As was described earlier\cite{MM.fmdcm:2022}, the essential quantity
to capture by a fluctuating distributed charge model are the charge
displacements $(u,v,w)$ along the local axes $\hat{\bf e}_{{\rm
    A},u}$, $\hat{\bf e}_{{\rm A},v}$, $\hat{\bf e}_{{\rm A},w}$ for
each charge $q$ associated with a reference atom A. Local axis systems
are invariant to molecule translation and rotation and approximately
retain the charge positions relative to selected neighboring atoms
upon conformational change.\cite{MM.dcm:2014} In contrast to MDCMs,
the charge displacements $(u(\mathbf{d}), v(\mathbf{d}),
w(\mathbf{d}))$ of kMDCMs depend on the geometry $\mathbf{R}$ of the
molecule, in particular the internal distances $\mathbf{d}$, and need
to be represented as smooth functions of the geometry. Within kMDCM
this is accomplished through solution of the associated kernel
equations.\\
  
For kMDCM the kernel function $K(\cdot,\cdot)$ predicts the 
charge displacements $\boldsymbol{\kappa}_q$ with components $\kappa^u_{q}$,
$\kappa^v_{q}$, $\kappa^w_{q}$ of a charge $q$ along one of the three
local axes as a sum of kernelized distances weighted by
$\boldsymbol{\alpha}$, according to
\begin{align}
    \kappa^u_q(\mathbf{d}') &= \sum^{N_\textrm{train}}_{i=1} \alpha^u_{q,i}
    K(\mathbf{d',d}_{i}) \label{eq:kernelu}\\
    \kappa^v_q(\mathbf{d}') &= \sum^{N_\textrm{train}}_{i=1} \alpha^v_{q,i}
    K(\mathbf{d',d}_{i}) \label{eq:kernelv}\\
    \kappa^w_q(\mathbf{d}') &= \sum^{N_\textrm{train}}_{i=1} \alpha^w_{q,i}
    K(\mathbf{d',d}_{i})
    \label{eq:kernelw}
\end{align}

The coefficients $\boldsymbol{\alpha}$ are obtained through the 
"kernel trick"\cite{gp06} 
by inverting the kernel matrix $\mathbf{K} = \{K(\mathbf{d_i}, \mathbf{d_j})\}_{i,j=1}^{N_\textrm{train}}$ to yield
\begin{align}
  \boldsymbol{\alpha}^u_{q} &= (\mathbf{K}
  +\lambda_2\mathbf{I})^{-1} \kappa^{u}_{q}
  \\ \boldsymbol{\alpha}^v_{q} &= (\mathbf{K}
  +\lambda_2\mathbf{I})^{-1} \kappa^{v}_{q}
  \\ \boldsymbol{\alpha}^w_{q} &= (\mathbf{K}
  +\lambda_2\mathbf{I})^{-1} \kappa^{w}_{q}.
  \label{eq:kerneltrick}
\end{align}
This equation was solved by using Cholesky\cite{Higham_2009}
decomposition for a subset of all conformations (i.e. the training or
reference set). An additional hyperparameter, $\lambda_2$, was added
to the diagonal of the kernel matrix, to regularize the problem with
respect to noise in the data and to stabilize the inversion for
near-singular kernel matrices. Adding the $\lambda_2-$weighted unit
matrix corresponds to introducing Gaussian-distributed noise with
variance $\lambda_2$.\cite{murphy2012machine} The efficiency of such
kernel-based methods scales with dataset size which may become
prohibitive.\\

The charge position $\mathbf{r}'_{q, {\rm A}}$ within the local axis
system of atom A and the position $\boldsymbol{\delta}'_{q, {\rm A}}$
in the global reference frame are determined by
\begin{align}
    \mathbf{r}'_{q,{\rm A}}(\mathbf{d}') &= 
    \kappa^u_{q}(\mathbf{d}') \hat{\bf e}_{{\rm A},u} 
    + \kappa^v_{q}(\mathbf{d}') \hat{\bf e}_{{\rm A},v} 
    + \kappa^w_{q}(\mathbf{d}') \hat{\bf e}_{{\rm A},w}
  \label{eq:global-displacement} \\
    \boldsymbol{\delta}'_{q}(\mathbf{d}') &= 
    \mathbf{r}'_{q,{\rm A}}(\mathbf{d}') + \mathbf{R}'_{\rm A}
  \label{eq:global-position} 
\end{align}
using the three local axis vectors $\hat{\bf e}_{{\rm A},u}$,
$\hat{\bf e}_{{\rm A},v}$, $\hat{\bf e}_{{\rm A},w}$ and the atom
position $\mathbf{R}'_{\rm A}$ of atom A. With the global positions
$\boldsymbol{\delta}'$ of all distributed charges, the ${\rm
  ESP}(\boldsymbol{\delta}', \mathbf{R}')$ can be computed. It is
useful to note that the charge position $\boldsymbol{\delta}_q$ for a
charge $q$ in Cartesian coordinates corresponds to displacements
$\boldsymbol{\kappa}_{q}$ in the local axis system and that
$\kappa^u_{q}(\mathbf{d})$ is the representation of
$u_{q}(\mathbf{d})$.\\

\begin{figure}[h!]
    \centering
    \includegraphics[width=0.8\textwidth]{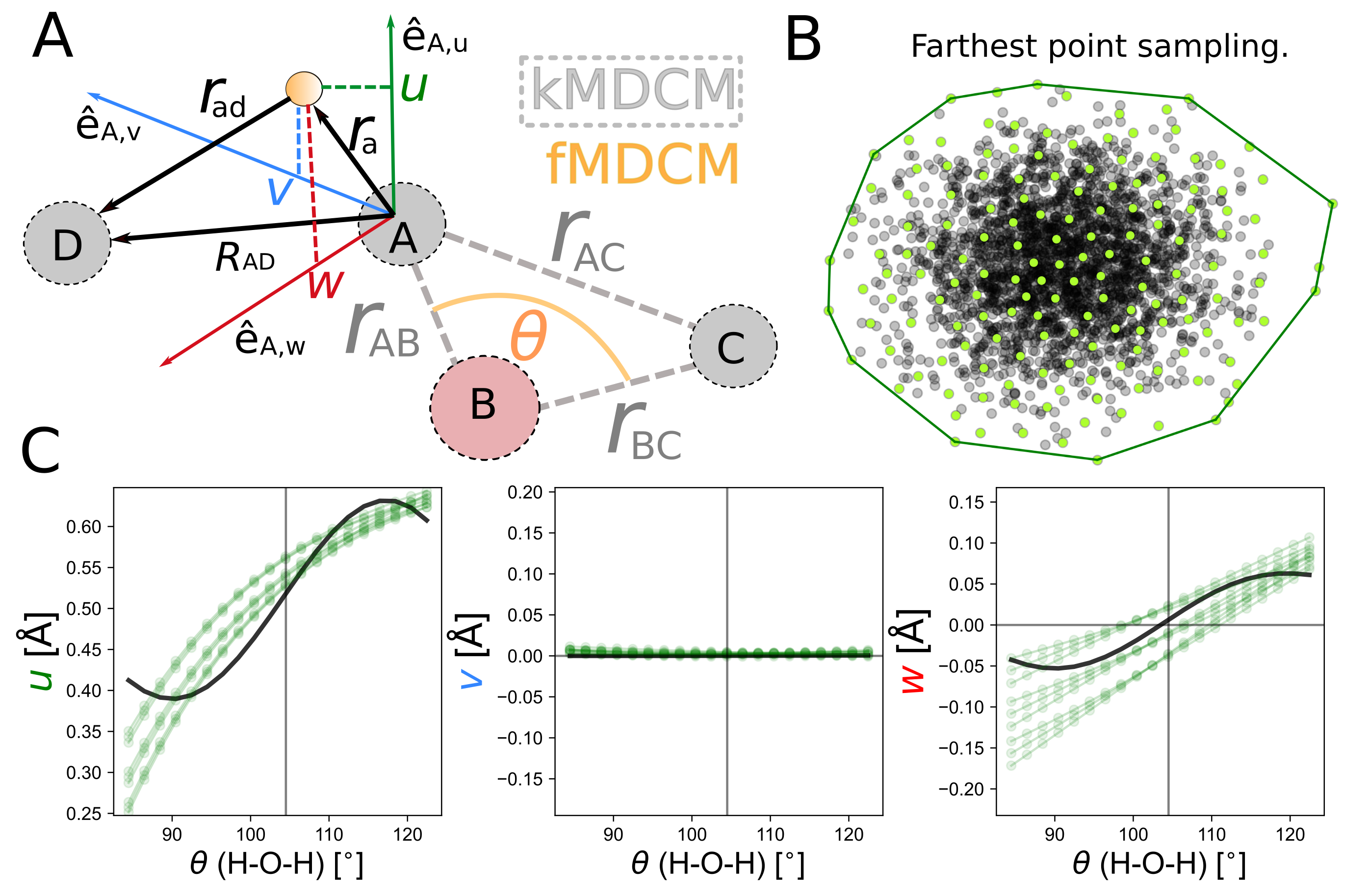}
    \caption{Panel A: Both fMDCM and kMDCM parametrize charge
      displacements (e.g. $r_a$ associated with atom A) along local
      axes $(u,v,w)$ using $(1-\cos{\theta}$) and internal distances
      $\mathbf{d}$, respectively.  Panel B: An illustration of
      farthest-point sampling in 2D is shown, improving coverage of
      conformational space. Kernels create a convex support based on
      the distance between data points.  Panel C: Comparison of the
      fitted local charge positions depending on $\theta$ for the
      kMDCM (green; multiple lines indicate different bond lengths)
      and fMDCM (black; $\theta-$dependent, polynomial-fitted charge
      displacements\cite{MM.fmdcm:2022}) for a single charge in a
      flexible six-charge water model. The vertical line is at the
      equilibrium angle $(\theta_{\rm e} = 104.5^{\circ})$; the
      horizontal line indicates 0 \AA\/ in local coordinates,
      corresponding to atom-centered charges.}
    \label{fig:kMDCM}
\end{figure}

\subsection{Kernel-Based MDCM derivatives}
For MD simulations the derivatives of the interactions with respect to
the Cartesian coordinates for each atom are required. As an example,
the Coulomb interaction $V$ between charge $q_{\rm a,kMDCM}$
associated with site A (treated with kMDCM) and $q_{\rm d,PC}$ at site
D (treated with atom-centered PC) is considered.  Omitting the prefactor 
$(4\pi \varepsilon_0)^{-1}$, the interaction is
\begin{equation}
  V \propto \frac{q_{\rm a,kMDCM} \cdot q_{\rm d,PC}}{|\mathbf{r}_{\rm ad}|}
\end{equation}
with $\mathbf{r}_{\rm ad} = 
 \mathbf{R}_{\rm D}
 - (\mathbf{r}_{\rm a} + \mathbf{R}_{\rm A})$,
see Figure \ref{fig:kMDCM}, which may depend on some
arbitrary molecular distortion $\mathbf{\rho}$, as described
previously\cite{MM.fmdcm:2022}.  With $\gamma = \{x,y,z\}$, the
derivative of the corresponding Coulomb potential with respect to some
change in position of atom A is
\begin{equation}
\frac{\partial V}{\partial R_{{\rm A},\gamma}} = q_{\rm a,kMDCM}
  \cdot q_{\rm d,PC} \frac{\partial }{\partial R_{{\rm A},\gamma}}
  \frac{1}{|{\mathbf{r}}_{\rm ad}|}
  \label{eq:DCM_dV}
\end{equation}

For any off-center, distributed
charge (DC) model - such as MDCM, fMDCM or kMDCM -
$\mathbf{r}_{\rm a}$ is defined relative to a local axis system
$\hat{\bf e}_{{\rm A},x}$, $\hat{\bf e}_{{\rm A},y}$, $\hat{\bf
  e}_{{\rm A},z}$ defined by atoms A, B and C, as described
elsewhere.\cite{MM.dcm:2014} Forces on an off-center charge $q_{\rm
  a,DC}$ associated with atom A generate torques on atoms A, B and C
according to Eq. \eqref{eq:DCM_dV} for these three atoms, replacing
$\partial R_{\rm A}$ by $\partial R_{\rm B}$ or $\partial R_{\rm C}$
as appropriate. The complete set of partial derivatives for
distributed charge $q_{\rm a,DC}$ is
\begin{align}
\frac{\partial V}{\partial R_{{\rm A},\gamma}} &= 
  -\frac{
    q_{\rm a,DC} q_{\rm d,PC} 
    \bigl(
      R_{{\rm AD},x}(\hat{\boldsymbol{\gamma}}\cdot\mathbf{\hat{x}} + g_{1\gamma})
      + R_{{\rm AD},y}(\hat{\boldsymbol{\gamma}}\cdot\mathbf{\hat{y}} + g_{2\gamma})
      + R_{{\rm AD},z}(\hat{\boldsymbol{\gamma}}\cdot\mathbf{\hat{z}} + g_{3\gamma})
    \bigr)}
    {|\boldsymbol{R}_{\rm AD}|^{3}}
  \label{Eq:DCMderiv1}\\ 
  \frac{\partial V}{\partial R_{{\rm B},\gamma}} &=
  -\frac{
    q_{\rm a,DC} q_{\rm d,PC}
    \bigl(
      R_{{\rm AD},x} g_{4\gamma} 
      + R_{{\rm AD},y} g_{5\gamma} 
      + R_{{\rm AD},z} g_{6\gamma}
    \bigr)}
    {|\boldsymbol{R}_{\rm AD}|^{3}}
  \label{Eq:DCMderiv2}\\ 
  \frac{\partial V}{\partial R_{{\rm C},\gamma}} &= 
  -\frac{
    q_{\rm a,DC} q_{\rm d,PC} 
    \bigl(
      R_{{\rm AD},x} g_{7\gamma} 
      + R_{{\rm AD},y} g_{8\gamma} 
      + R_{{\rm AD},z}g_{9\gamma}
    \bigr)}
    {|\boldsymbol{R}_{\rm AD}|^{3}}
\label{Eq:DCMderiv3}
\end{align}
where the scalar product
$\hat{\boldsymbol{\gamma}}\cdot\mathbf{\hat{x}}$ is 1 for $\gamma=x$
and zero otherwise, $R_{{\rm AD},x}$ is the $x$-component of vector
$\mathbf{R}_{\rm AD}$ pointing from atom A to D and the
coefficients $g_{1\gamma}$ to $g_{9\gamma}$ contain the partial
derivatives of the local unit vectors of the frame
($\mathbf{\hat{e}}_{u}, \mathbf{\hat{e}}_{v}, \mathbf{\hat{e}}_{w}$)
with respect to the nuclear coordinate components $R_{{\rm
    A},\gamma}$, $R_{{\rm B},\gamma}$ and $R_{{\rm
    C},\gamma}$.\cite{MM.dcm:2014} \\
    
\begin{figure}[h!]
    \centering
    \includegraphics[width=0.8\textwidth]{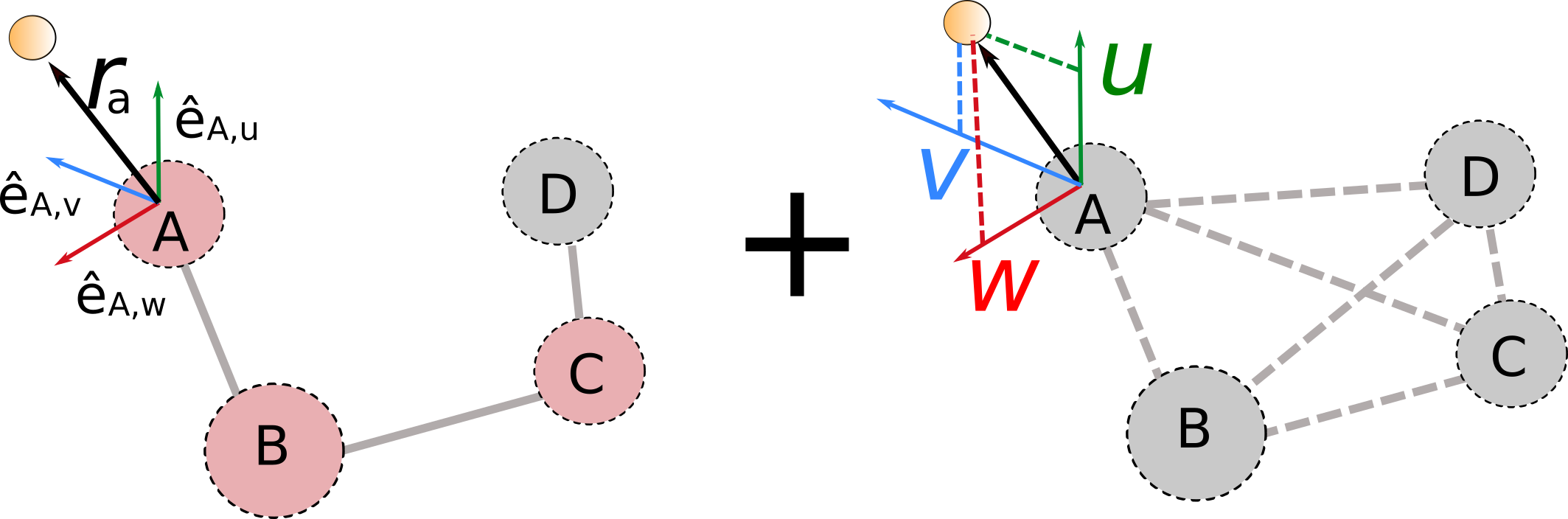}
    \caption{For an example tetra-atomic molecule, the charge
      displacement vector $\mathbf{r}_{a}$ depends on the local
      axes defined by the atoms (A,B,C), and the charge displacements
      along the local axes $u,v,w$.  Eq. \ref{Eq:cderiv2} defines the
      change in the $x$ component of $\mathbf{r}_{a}$ given
      displacement of atom $A$ in the $x$ direction.}
    \label{fig:enter-label}
\end{figure}
    
Since $\kappa^u_{q}$, $\kappa^v_{q}$, and $\kappa^w_{q}$ depend on
the interatomic distances vector
$\mathbf{d}'$, partial
derivatives of the local unit vectors with respect to the
nuclear coordinates contain contributions from each of the
intermolecular distances involving atom A. 
As an example, the partial derivative $g_{1\gamma}(\mathbf{d'})$ for the
$x-$component of the Cartesian derivative of atom A is
\begin{align}
  g_{1\gamma}(\mathbf{d'}) = &  
    \Bigg[
      \kappa^u_{q}(\mathbf{d}')
      \frac{\partial ({\hat{\bf e}}_{{\rm A},u})_{x}}{\partial R_{{\rm A},\gamma}}
      + ({\hat{\bf e}}_{{\rm A},u})_{x} 
               \frac{\partial \kappa^u_{q}(\mathbf{d}')}{\partial R_{{\rm A},\gamma}}
    \Bigg] + \nonumber \\ 
    & 
    \Bigg[ 
      \kappa^v_{q}(\mathbf{d}')
      \frac{\partial ({\hat{\bf e}}_{\rm A},v)_{x}}{\partial R_{{\rm A},\gamma}} 
      + ({\hat{\bf e}}_{{\rm A},v})_{x} 
        \frac{\partial \kappa^v_{q}(\mathbf{d'})}{\partial R_{{\rm A},\gamma}}
    \Bigg] + \nonumber \\ 
    & 
    \Bigg[
       \kappa^w_{q}(\mathbf{d}')
       \frac{\partial ({\hat{\bf e}}_{{\rm A},w})_{x}}{\partial R_{{\rm A},\gamma}} 
       + ({\hat{\bf e}}_{{\rm A},w})_{x} 
         \frac{\partial \kappa^w_{q}(\mathbf{d'})}{\partial R_{{\rm A},\gamma}}
    \Bigg] ~.
 \label{Eq:cderiv2}
\end{align}

Derivatives of the type $\partial \kappa^u_{q}(\mathbf{d}') / \partial
R_{{\rm A},\gamma}$ are evaluated using the chain rule, summing over
all $N_{a} - 1$ intramolecular distances with index $k$ including atom
A to the remaining atoms X$\neq$A
\begin{equation}
\frac{\partial \kappa^u_{q}(\mathbf{d}')}{\partial R_{{\rm A},\gamma}}
= \sum^{N_{\mathrm{a}} - 1}_{k=1} \frac{\partial
  \kappa^u_{q}(\mathbf{d}')}{\partial d'_k} \frac{\partial
  d'_k}{\partial R_{{\rm A},\gamma}}
\end{equation}
where
\begin{align}
    \frac{\partial \kappa^u_{q}(\mathbf{d}')}{\partial d'_k} =& 
    - \sum^{N_{\rm train}}_{i=1} \alpha_{i} K(\mathbf{d',d}_{i})
    \frac{(d'_{k} - d_{i,k})}{\sigma^2}
    \label{eq:grad1} \\
    \frac{\partial d'_k}{\partial R_{{\rm A},\gamma}} =&
    \frac{R_{{\rm A},\gamma} - R_{{\rm X},\gamma}}{d_{k}}
    \label{eq:grad2}
\end{align}
and $d_k = | \mathbf{R}_{\rm A} - \mathbf{R}_{\rm X} |$ are internal
distances involving atom A and X$\neq$A.  This completes the
derivation of analytical forces in Eq. \ref{eq:DCM_dV}.\\

\noindent
As fMDCM also attempts to describe intramolecular charge
redistribution, a model for water was constructed to compare directly
with kMDCM presented in the present work. As an improvement over the
earlier parametrization\cite{MM.fmdcm:2022} which used a third order
polynomial $\sum_{i=0}^{3} a_i \theta^i$ to describe the coordinate
dependence of $(u(\theta),v(\theta),w(\theta))$, an expansion
$\sum_{i=0}^{3} a_i (1-cos(\theta))^i$ was used in the following. This
also improves the quality of the model when evaluated at angles close
to 180$^{\circ}$ which is particularly relevant for linear molecules
such as SCN$^-$.\cite{Toepfer:2022}.\\

\subsection{\textit{Ab initio} Reference Data}
The molecular ESP, obtained from the converged SCF density at the
PBE0/aug-cc-pVDZ level of theory was analysed using the CubeGen
utility in Gaussian16\cite{g16} with a grid spacing of 1.67
points/\AA\/.  Water geometries were obtained by enumerating a
$Z$-matrix representation on a 3-dimensional grid containing ten
evenly spaced angles between $84.45^\circ$ to $120.45^{\circ}$, and
each OH bond length sampled at 0.909 \AA\/, 0.959 \AA\/ and 1.009
\AA\/, resulting in 180 structures.  For methanol, 2500 structures
were sampled from gas phase MD simulations, using CHARMM\cite{charmm}
and CGenFF\cite{cgenff} parameters, with bond lengths involving
hydrogen atoms constrained using the SHAKE
algorithm.\cite{Ryckaert:1977} For generating methanol geometries, gas
phase simulations were performed at 600 K using an integration time
step $\Delta t = 1$ fs. First, temperature was equilibrated using the
Nosé-Hoover thermostat over 500 ps of simulation in the $NVT$
ensemble. 2.5 ns of simulation was then performed in the $NVE$
ensemble, with structures saved every 1 ps.\\

\noindent
As with all machine learning-based methods, samples to train the model
are required. Kernel models such as kMDCM use atom-atom distances
within the input space as a basis for a prediction. As such, data
coverage is an important consideration to ensure all relevant
conformations are adequately described. Farthest-point sampling was
used to select the conformations included in the training set to
maximize the diversity of the structures used for training. This was
achieved by randomly selecting an initial geometry and choosing each
subsequent geometry by selecting the one with the largest Euclidean
distance from the previous selection. An example of this procedure in
2D is illustrated in Figure \ref{fig:kMDCM}B, where training examples
at the boundary of the data distribution form the convex hull of the
kernel. \\

\subsection{MD Simulations using kMDCM}
All molecular dynamics (MD) simulations were run using the CHARMM 
package.\cite{charmm} Routines for calculating electrostatic energies
and forces for kMDCM were implemented inside the DCM
module.\cite{MM.mdcm:2017} To assess the validity and energy
conservation of the kMDCM implementation, a periodic water box with a
side length of 41 \AA\/ was set up containing 2000 water
molecules. For the intramolecular potential of water, 
an RKHS\cite{MM.rkhs:2017} representation of reference 
energies at the PNO-LCCSD(T)/aug-cc-pVTZ level of theory 
was used.\cite{MM.hydra:2023}\\

For equilibration, the system was simulated in the $NpT$ ensemble at
$T = 300$ K and $p = 1$ atm for 1 ns using $\Delta t = 0.2$ fs with
van der Waals parameters taken from CGenFF.\cite{cgenff} Nonbonded
interactions were truncated at 16 \AA\/ and electrostatic interactions
were calculated using particle mesh Ewald summation. After
equilibration the volume was kept constant during 500 ps of simulation
in $NVT$, after which the thermostat was turned off and a further 1 ns
of simulation in the $NVE$ ensemble was performed for data
accumulation. 
Additionally, IR spectra were computed from the simulation data. 
The IR line shape, $I(\omega)$, was obtained via the
Fourier transform of the dipole–dipole autocorrelation function from
the dipole moment time series, with snapshots saved every 2 fs from a
500 ps $NVE-$simulation. For comparison, the anharmonic frequencies
for a single gas phase water molecule calculated using
DVR3D\cite{Tennyson:2004} for the bend, symmetric and asymmetric
stretch were 1575, 3690, and 3719 cm$^{-1}$,
respectively,\cite{MM.hydra:2023} compared with experimentally
observed frequencies\cite{shimanouchi:1972} at 1595, 3657, and 3756
cm$^{-1}$ which validates the bonded terms. \\

\section{Results}
\subsection{Quality of the ESP}
\begin{figure}[h!]
    \centering
    \includegraphics[width=0.5\textwidth]{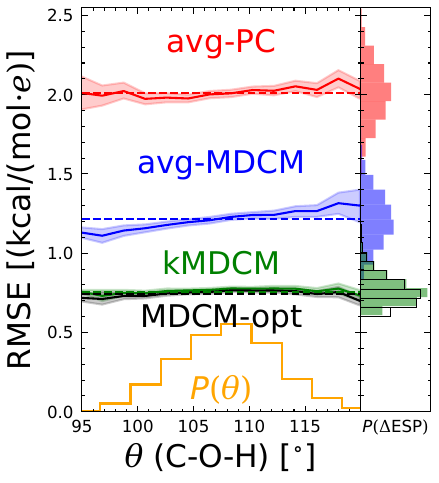}
    \caption{Left: The average ESP quality for methanol depending on
      the COH angle with static charge models avg-PC (red) and
      avg-MDCM (blue) compared with conformationally dependent kMDCM
      (green) and optimized positions using l-BFGS (black), evaluated
      on the test set. The shaded area shows the 99\% confidence
      interval and the dashed horizontal lines indicate the position
      of the median RMSE.  Right: the normalized distributions for
      $P(\Delta{\rm ESP}(\theta))$. The two static models exhibit
      conformational dependence w.r.t the COH angle. The distribution
      of angles is shown in orange.}
    \label{fig:methanol}
\end{figure}

First, methanol based on a 10 distributed charge model
was selected as a test system to
validate kMDCM and to assess the conformational dependence of the
molecular ESP upon changes in angles and dihedrals. For this, 2500
methanol conformations were generated from a gas phase MD
simulation initialized from velocities corresponding to a Maxwell-Boltzmann
distribution at 300 K. For the kMDCM model 256 structures were used
for training and the average RMSE for the test set was 1.0
kcal/(mol$\cdot e)$. When fitting to the remaining structures
using conformational averaging, the RMSE was 1.2 kcal/(mol$\cdot e)$
for static MDCM, which increased to 2.0 kcal/(mol$\cdot e)$ for an
atom-centered PC model, see Figure \ref{fig:methanol}.
Additionally, a dependence of the quality of the ESP on the COH angle
was observed. Although averaging over an ensemble of structures is
recommended\cite{Jensen:2022, Jensen:2017, Koch:1995} this may lead to
implicit bias in the quality of the models for certain conformations,
as seen in Figure \ref{fig:methanol} (blue line).  Distorted
structures with COH angles $\theta \leq100^{\circ}$ yielded smaller
errors in comparison to more linear structures ($\theta \geq
160^\circ$) by approximately 0.2 kcal/(mol$\cdot e)$. The
conformational dependence is less pronounced for PC models, likely due
to the lower complexity of the fitting parameters and the higher
baseline errors. The width of the error distribution of the new kMDCM
model is smallest and no bias towards particular angles is observed.\\

Next, the performance of kMDCM for a water molecule was assessed by a
systematic scan ($3\times3\times20 = 180$ structures for
$r_{\mathrm{OH_1}}$, $r_{\mathrm{OH_2}}$, and
$\theta_{\mathrm{H_1OH_2}}$) along both OH-bonds and the valence angle
for water geometries.  A 6 charge kMDCM model from farthest-point
sampling using 16 structures was generated with an additional 164
structures available as the test set. The trained model achieved an
average RMSE of 0.7 kcal/(mol$\cdot e)$ for the test set with a
maximum error of 0.8 kcal/(mol$\cdot e)$. This compares with an
average RMSE and maximum error of 0.7 kcal/(mol$\cdot e)$ and 1.1
kcal/(mol$\cdot e)$, respectively, for a fMDCM model fit to all
structures. The favourable performance of the kMDCM and fMDCM models
can be related to the fact that OH bond-coupling, which fMDCM is
insensitive to, has less impact on the ESP changes than the variation
of the valence angle.  For the ensemble averaged MDCM, the water
conformer with the highest RMSE was equal to 1.8 kcal/(mol$\cdot e)$,
and was 1.0 kcal/(mol$\cdot e)$ on average for the entire
distribution. In comparison, optimizing the ESP of a PC model to an
average of all structures, gave an average RMSE of 2.3 kcal/(mol$\cdot
e)$ and a maximum RMSE of 2.7 kcal/(mol$\cdot e)$.  For water monomers
with non-equilibrium bond lengths which incurred the highest errors,
the improvement of $\sim30\%$ for the RMSE between kMDCM and fMDCM
shows the advantage of incorporating all conformational degrees of
freedom into the kernel-based model. For both, fMDCM and kMDCM, the
representation of $(u,v,w)$ depends in a similar fashion on the
valance angle $\theta$, see Figure \ref{fig:kMDCM}C. However, kMDCM
which depends on all internal degrees of freedom captures this
dependence in a more realistic fashion than fMDCM for which a
third-order polynomial is not sufficiently flexible to describe the
variation of $(u,v,w)$ with $\theta$ and does not depend on OH bond
length.\\

\subsection{Molecular Dipole Surface}
The electrostatic models in this study were fit to reproduce the 
reference molecular ESP. However, modelling the conformational
dependence of the molecular molecular dipole moment is equally
important in computational spectroscopy as the intensity of IR
transitions depend on it. To determine how the molecular dipole moment
of methanol changes with conformation, simulations of a single
methanol with constrained bonds involving hydrogen atoms were
performed, and snapshots were taken every $0.1$ fs. The magnitude of
the molecular dipole moment ${\mu}$ was computed for 2500 snapshots
using the kMDCM and fMDCM models described above and compared with
{\it ab initio} values calculated at the PBE0/aug-cc-pVDZ level of
theory used for the reference calculations.\\

\begin{figure}
    \centering
    \includegraphics[width=1.\textwidth]{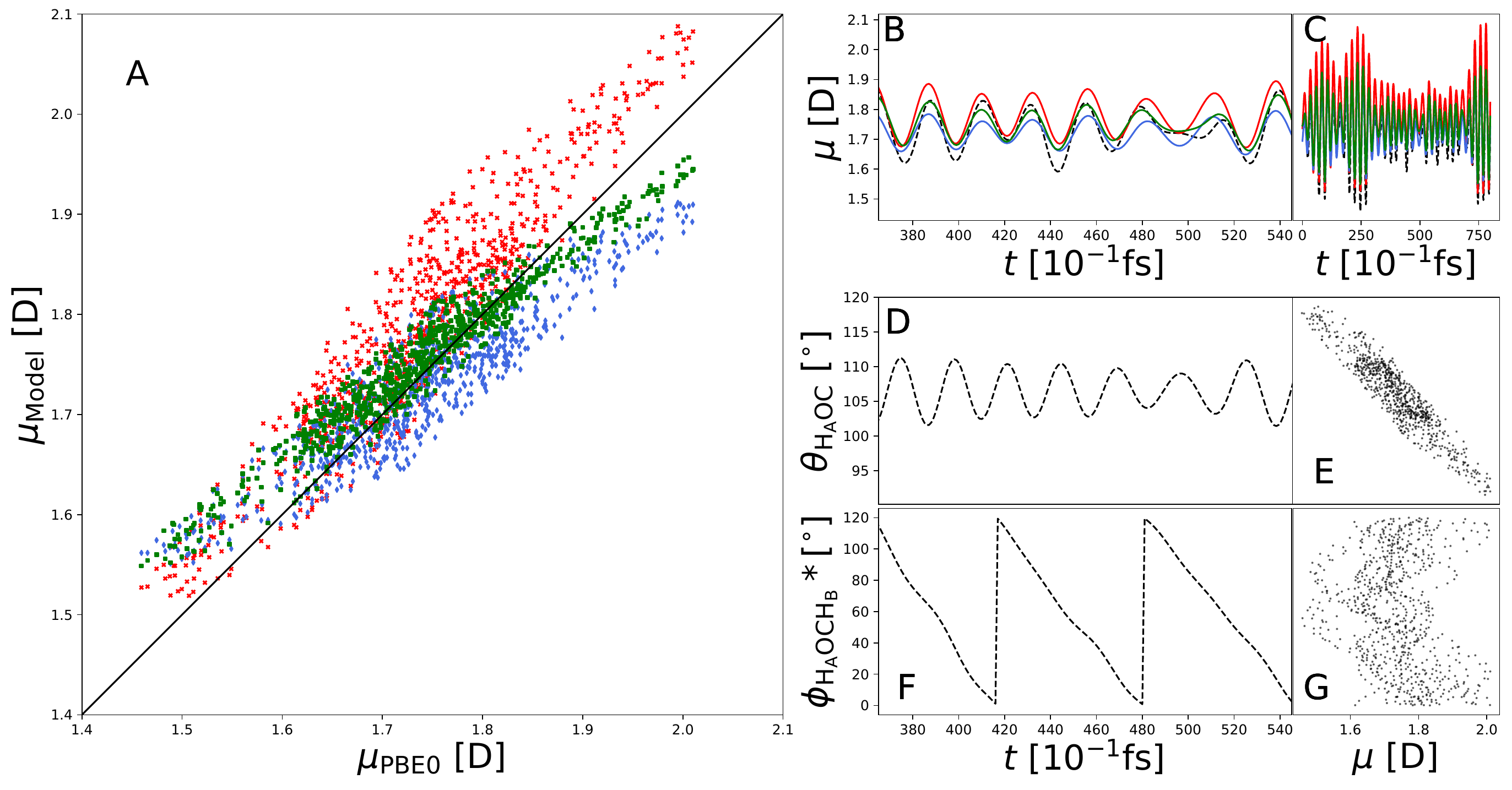}
    \caption{Performance of kMDCM, MDCM and PC models for methanol
      from MD-generated structures with constrained bonds involving
      hydrogen atoms. Panel A: The magnitude of the molecular dipole
      moment from DFT calculations $\mu_{\rm PBE0}$ versus $\mu_{\rm
        PC}$ (red), $\mu_{\rm MDCM}$ (blue), and $\mu_{\rm kMDCM}$
      (green).  Panels B and C: Model and reference (black dashed
      line) dipole moment time series $\mu(t)$ (zoom-in and full
      sequence); for color code see panel A. Panel D: Corresponding
      time series of the H$_{\mathrm{A}}$OC angle. Panel E:
      Correlation between H$_{\rm A}$OC angle and the reference
      $\mu_{\rm PBE0}$ Panel F: H$_{\mathrm{A}}$OCH$_{\mathrm{B}}$
      dihedral time series ($^{*}$modulo 120 to account for the three
      fold symmetry of the rotation). Panel G: The reference dipole
      moment $\mu_{\rm PBE0}$ versus
      H$_{\mathrm{A}}$OCH$_{\mathrm{B}}$.}
    \label{fig:dipoles}
\end{figure}

A direct comparison of the molecular dipole determined from different
models for the ESP with the reference electronic structure
calculations is given in Figure \ref{fig:dipoles}A. Depending on the
conformation, the molecular dipole from reference PBE0 calculations
$\mu_{\rm PBE0}$ varies between 1.5 and 2.0 D which demonstrates that
the conformational dependence needs to be taken into account. The
kMDCM model (green) yielded $\mu_{\rm kMDCM}$ with an RMSE of 0.04 D
compared with $\mu_{\rm PBE0}$ and was effective at describing the
conformationally dependent molecular dipole moment. This compares with
RMSEs of $0.07$ D and $0.05$ D for $\mu_{\rm PC}$ and $\mu_{\rm MDCM}$
from conformationally independent PC and MDCM models (Figure
\ref{fig:dipoles}A).\\

\noindent
It is also interesting to consider the time-dependence of $\mu(t)$
(panels B and C) explicitly and to investigate whether its variation
is correlated with particular internal motions, see panels D to G in
Figure \ref{fig:dipoles}. The two coordinates considered were the
H$_{\rm A}$OC valence angle and the H$_{\rm A}$COH$_{\rm B}$ dihedral
(where H$_{\rm B}$ are the methyl hydrogen atoms). Comparing panels D
(angle) and E (dipole) clarifies that the variation of $\mu_{\rm
  PBE0}(t)$ correlates with changes in the valence angle whereas the
dihedral motion does not correlate directly with the dipole moment,
see panels F (dihedral) and G (dipole). \\

\subsection{Impact of Model Parameters and Iterative Refinement}
Besides the number of charges considered, a kMDCM model requires two
hyperparameters to be chosen: $\lambda_1$ and $\lambda_2$. These may
depend on the molecule considered and the purpose for which the model
is developed. The impact of these hyperparameters is discussed for
kMDCM models for water (6 charges each with $u$, $v$, $w$ to yield 18
kernels and 3 atom-atom separations) and for methanol (10 charges, 30
kernels, 15 atom-atom separations). As per the methods section,
$\lambda_1$ adds a penalty to the loss function which increases with
the displacement of charges with respect to their initial
positions. For water, for which the loss landscape was found to be
smoother compared with that for methanol, $\lambda_1 = 0$
kcal/(mol$\cdot e\cdot$\AA) was found to be effective. On the other
hand, for methanol, $\lambda_1 = 50$ kcal/(mol$\cdot e\cdot$\AA) was
used initially to prevent charges from moving too far and to better
control and guide the optimization.\\

\begin{figure}[h!]
    \centering
    \includegraphics[width=0.99\textwidth]{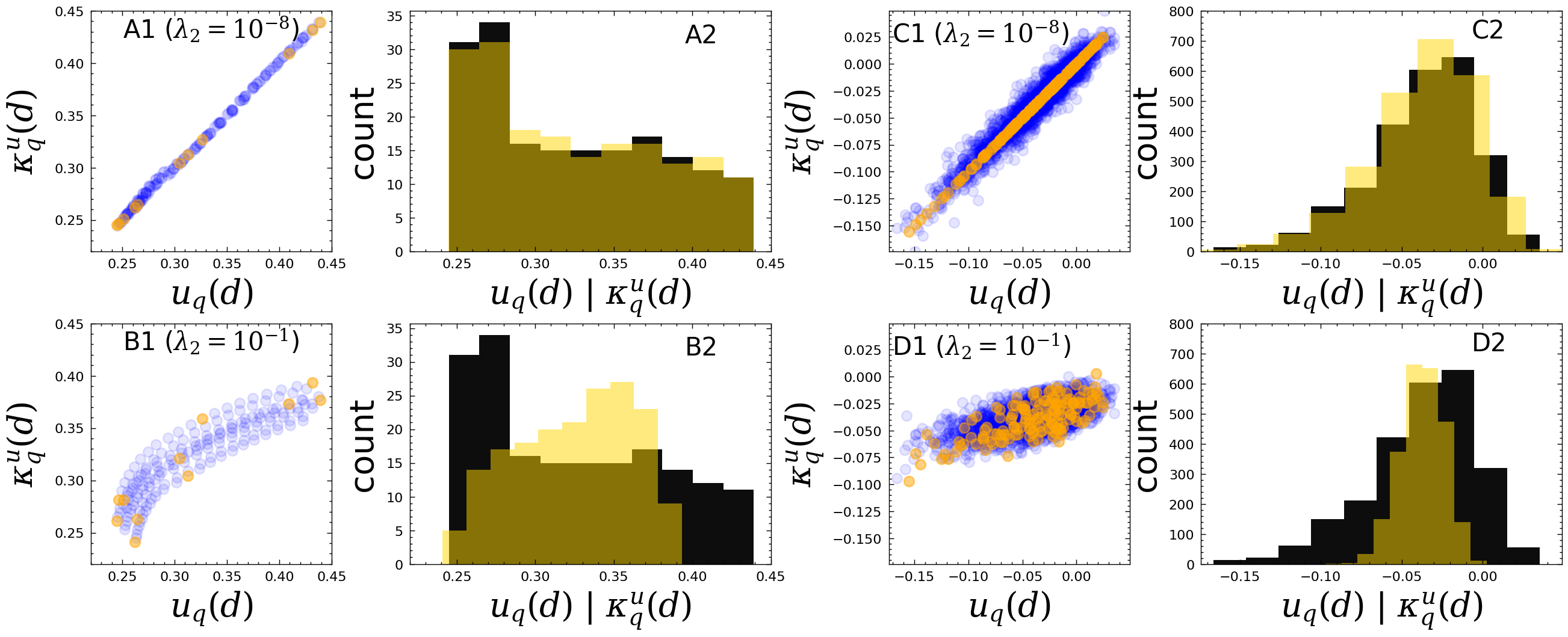}
    \caption{Correlation between optimized local displacement
      components ($u_q(d)$) and corresponding kernel predictions
      ($\kappa_q^u(d)$) for different values of the kernel
      regularization parameter ($\lambda_2$) for water (A1, B1) and
      methanol (C1, D1). Test and train examples are depicted as blue
      and orange circles, respectively. To the right, the accompanying
      distributions of $u_q(d)$ (black) and $\kappa_q^u(d)$ (yellow)
      the test set are shown (A2-D2).}
    \label{fig:si-ku}
\end{figure}

\begin{table}[h!]
    \centering
    \begin{tabular}{c||c|c||c|c}
        \textbf{$\lambda_2$} & \textbf{$\langle \mathrm{RMSE} \rangle^{\mathrm{dis.}}_\mathrm{test}$} & \textbf{$\langle \mathrm{RMSE} \rangle^{\mathrm{ESP}}_\mathrm{test}$} & \textbf{$\langle \mathrm{RMSE} \rangle^{\mathrm{dis.}}_\mathrm{train}$} & \textbf{$\langle \mathrm{RMSE} \rangle^{\mathrm{ESP}}_\mathrm{train}$} \\
        \hline
        $10^{-8}$ & 0.006 & 0.60 & 0.000 & 0.62 \\
        $10^{-3}$ & 0.014 & 0.61 & 0.012 & 0.63 \\
        $10^{-1}$ & 0.036 & 0.75 & 0.038 & 0.79 \\
        $10^0$ & 0.078 & 0.96 & 0.084 & 1.03 \\
        \hline
    \end{tabular}
    \caption{The average displacement between optimised and predicted
      charge positions ($\langle \mathrm{RMSE} \rangle^{\mathrm{dis.}}$) 
      for the kMDCM
      model of water in units of \AA. The average RMSE between ESPs of
      optimized and predicted charge positions
      ($\langle \mathrm{RMSE} \rangle^{\mathrm{ESP}}$) in kcal/(mol$\cdot e$).}
    \label{tab:alpha_water}
\end{table}

\begin{table}[h!]
    \centering
    \begin{tabular}{c||c|c||c|c}
        \textbf{$\lambda_2$} & \textbf{$\langle \mathrm{RMSE} \rangle^{\mathrm{dis.}}_\mathrm{test}$} & \textbf{$\langle \mathrm{RMSE} \rangle^{\mathrm{ESP}}_\mathrm{test}$} & \textbf{$\langle \mathrm{RMSE} \rangle^{\mathrm{dis.}}_\mathrm{train}$} & \textbf{$\langle \mathrm{RMSE} \rangle^{\mathrm{ESP}}_\mathrm{train}$} \\
        \hline
        $10^{-8}$ & 0.004 & 1.07 & 0.000 & 1.07 \\
        $10^{-3}$ & 0.007 & 1.07 & 0.004 & 1.08 \\
        $10^{-1}$ & 0.016 & 1.11 & 0.016 & 1.11 \\
        $10^0$ & 0.023 & 1.15 & 0.026 & 1.15 \\
        \hline
    \end{tabular}
    \caption{The average displacement between optimised and predicted
      charge positions ($\langle \mathrm{RMSE} \rangle^{\mathrm{dis.}}$) for the kMDCM
      model of methanol in units of \AA. The average RMSE between ESPs of
      optimized and predicted charge positions
      ($\langle \mathrm{RMSE} \rangle^{\mathrm{ESP}}$) in kcal/(mol$\cdot e$).}
    \label{tab:alpha_methanol}
\end{table}

\noindent
Secondly, the kernel regularization parameter $\lambda_2$ weighs the
trade-off between over- and under-fitting.  For $\lambda_2 = 10^{-8}$
(Figure \ref{fig:si-ku} A.1 and C.1) training examples (orange
circles) are reproduced exactly. For water with only 3 degrees of
freedom the test examples (blue circles) can be interpolated given the
training examples, with $\langle {\rm RMSE}
\rangle^{\mathrm{dis.}}_{\rm test} = 0.006$ \AA\/ between the 
reference values $(u,v,w)$ after l-BFGS optimization and their representations ($\kappa_q^u$,
$\kappa_q^v$, $\kappa_q^w$) with $\lambda_2 = 10^{-8}$. Furthermore,
performance on the test set in terms of the average RMSE of the ESP
across all conformers is lowest (Table \ref{tab:alpha_water}). 
It is also noted that the
range of the predictions increases for smaller values of $\lambda_2$
(Figure \ref{fig:si-ku}B.1, yellow histogram) compared with the
reference values (black histogram).
Conversely, for larger values of $\lambda_2$, the range of predicted
values becomes smaller than that of the training examples (Figure
\ref{fig:si-ku} B.2 and D.2, yellow histograms), as
$\lambda_2$ increases the effective cut-off distance to training
examples in the input space causing predictions to be `averaged'
across more training structures. As a consequence, the distributions
$P(\kappa^u_{q})$, $P(\kappa^v_{q})$, and $P(\kappa^w_{q})$ become
more closely centered around the mean. For the high dimensional case,
methanol, $\lambda_2 = 10^{-3}$ was found to achieve similar performance in the test set when compared to $\lambda_2 = 10^{-8}$; however, the average distance between ground-truth
and predicted charge positions from 0.004 \AA\/ with low regularization
to 0.007 \AA~ (Table \ref{tab:alpha_methanol}). \\

\begin{figure}[h!]
    \centering
    \includegraphics[width=0.7\textwidth]{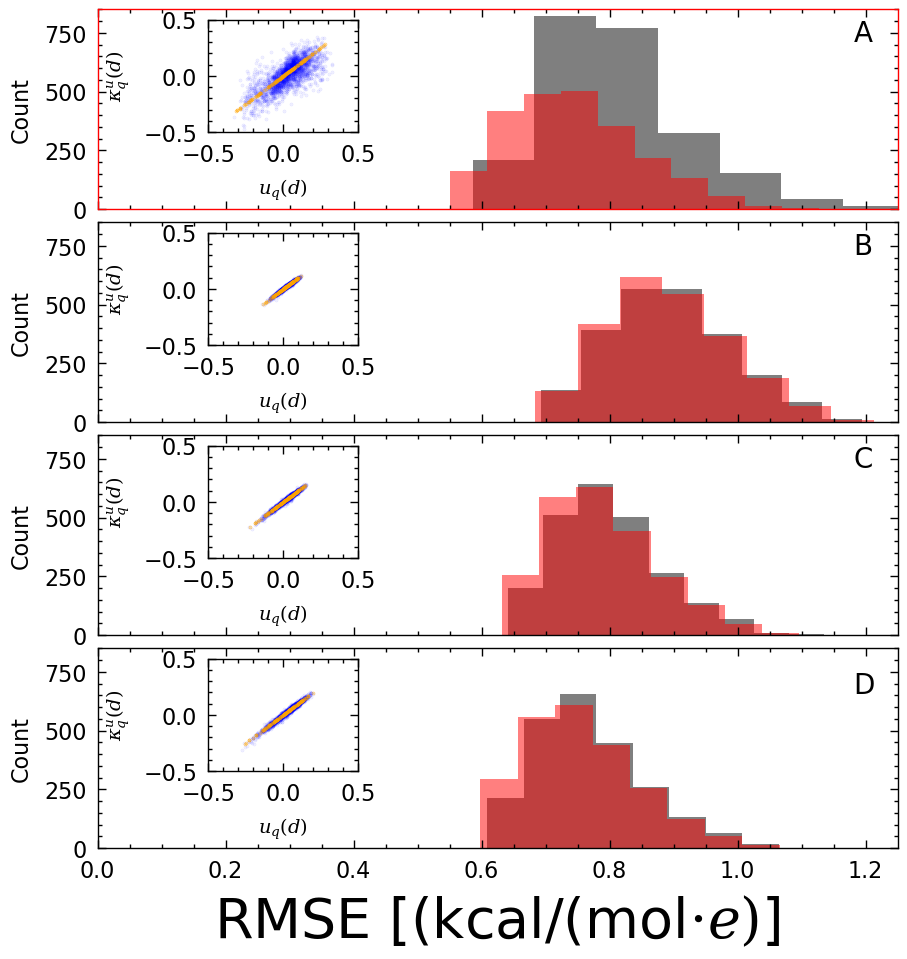}
    \caption{Distributions of
      $\mathrm{RMSE}^{\mathrm{ESP}}_\mathrm{test}$ for l-BFGS
      optimized charge positions (red) versus kMDCM (black), for (A) no
      $\lambda_1$ penalty, and (B, C, D) iterative refinement after 1,
      3, and 5 steps, respectively, using $\lambda_1 = 10$
      kcal/(mol$\cdot e \cdot$\AA). Insets show the correlation
      between the $u_q(d)$ and the kernel predictions,
      $\kappa_q^u(d)$. Kernels were fit with $\lambda_2 =10^{-8}$.}
    \label{fig:optimization}
\end{figure}

To generate ensembles of charge positions that are amenable to
interpolation using kMDCM, an iterative refinement scheme for fitting
the kMDCMs was developed.  Without adding a penalty the the charge
displacements ($\lambda_1 = 0$ kcal/(mol$\cdot e\cdot$\AA)) and
starting from a conformationally averaged MDCM model for methanol, it
was observed that the charge displacements were poorly interpolated
(with $\lambda_2 =10^{-8}$) using the kernel (Figure
\ref{fig:optimization}A, inset). The average and the range of
$\mathrm{RMSE}^{\mathrm{ESP}}_\mathrm{test}$ was larger in comparison
to the l-BFGS optimized models (Figure \ref{fig:optimization}A), which
may be related to the optimizer finding discontinuous local minima
when no charge displacement penalty is applied, leading to inaccurate
predictions.  When a penalty of $\lambda_1 = 50$ kcal/(mol$\cdot
e\cdot$\AA) was used to restrict the displacement of the charges, the
range of the charge displacements obtained by l-BFGS was small and
reproduced accurately by the kernel \ref{fig:optimization}B.  This
process was repeated by initializing the charge positions from the
kernel prediction allowed the range of the predicted charge positions
to increase in a controlled fashion, see Figure
\ref{fig:optimization}B-D. Following this iterative refinement lowered
the RMSE for the test set while keeping the accuracy of the kernel's
predictions stable during the iterative refinement.\\

\subsection{Molecular Dynamics Simulations with kMDCM}
Next, the feasibility of meaningful MD simulations using kMDCM was
assessed. This required that energy-conserving trajectories can be
generated from simulations in the $NVE$ ensemble. For this, the
formalism presented in the Methods section was implemented into the
DCM module in CHARMM version c48. MD simulations were then run with
2000 water molecules in an equilibrated periodic box at 1 atm pressure
without SHAKE constraints.  The electrostatic model was a 6 charge
kMDCM model, using Lennard-Jones parameters from the TIP3P water
model\cite{Jorgensen:1983} and the bonded terms were those of a RKHS
representation,\cite{MM.rkhs:2017} see Methods section. As the
electrostatic contribution of the force field was altered from the
original parametrization, further refinement of non-bonded parameters
will be necessary, e.g. along previously suggested
lines,\cite{MM.ff:2024} for direct comparison of computed observables
with experimentally measured properties. Such refinement was, however,
not deemed necessary for validating the implementation and was not
further pursued here.\\

\begin{figure}[h]
    \centering
    \includegraphics[width=0.80\textwidth]{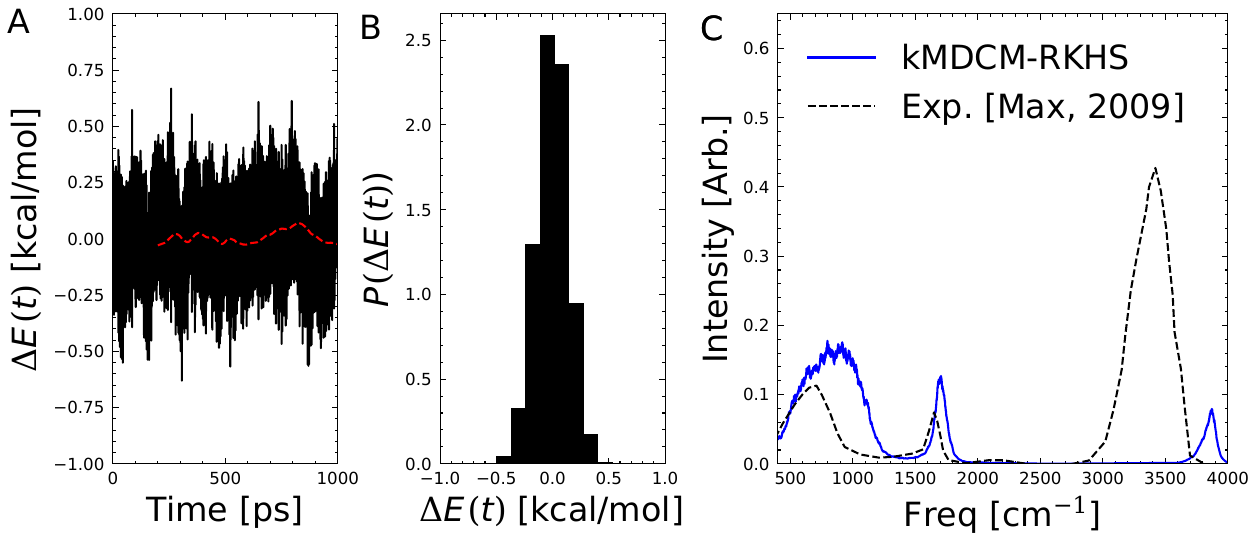}
    \caption{(A) Fluctuations from the average total energy, $\Delta
      E(t) = E(t) - \langle E \rangle$, from an $NVE$ simulation of a
      2000 molecule water box, using a time step of 0.2 fs. with
      kMDCM+RKHS. A running average over 200 ps is shown as a red,
      dashed line, (B) the corresponding distribution. (C) The
      simulated liquid phase IR spectrum in comparison with
      experiment\cite{Chapados:2009}. The intensity of the spectra was
      post-processed with a moving average over a sliding window of 15
      cm$^{-1}$.}
    \label{fig:MD}
\end{figure}

The combination of RKHS and kMDCM was found to be stable over the 1 ns
long $NVE$ simulation, based on the variance in total energy over time
and the corresponding distribution (Figures \ref{fig:MD}A and
B). Simulations using kMDCM exhibit variances in the total energy of
$\pm 0.5$ kcal/mol and $\pm 0.1$ kcal/mol for $\delta t=0.2$ fs and
$\delta t=0.1$ fs, respectively. For reference, rigid TIP3P using
SHAKE constraints leads to a variance in the total energy of $\pm 0.5$
kcal/mol using an integration time step of 1 fs.\cite{MM.fmdcm:2022}
As a test of the bonded terms, the simulated IR spectrum reproduces
the low frequency features (Figure \ref{fig:MD}C). The frequency of
the OH stretch vibrations are blue shifted in comparison with
experiment which is known for high-frequency modes from MD simulations
at 300 K.\cite{suhm:2020} Shifts in the low frequency modes may be
related to the choice of non-bonded parameters, as the system had a
density of 1.1 g/cm$^3$ which is 10\% larger than the experimental
value. Improvements to the IR and other observables can be achieved
through appropriate parametrization of the Lennard-Jones parameters
which is outside the scope of the present work.\\

\begin{figure}
    \centering
    \includegraphics[width=1.0\textwidth]{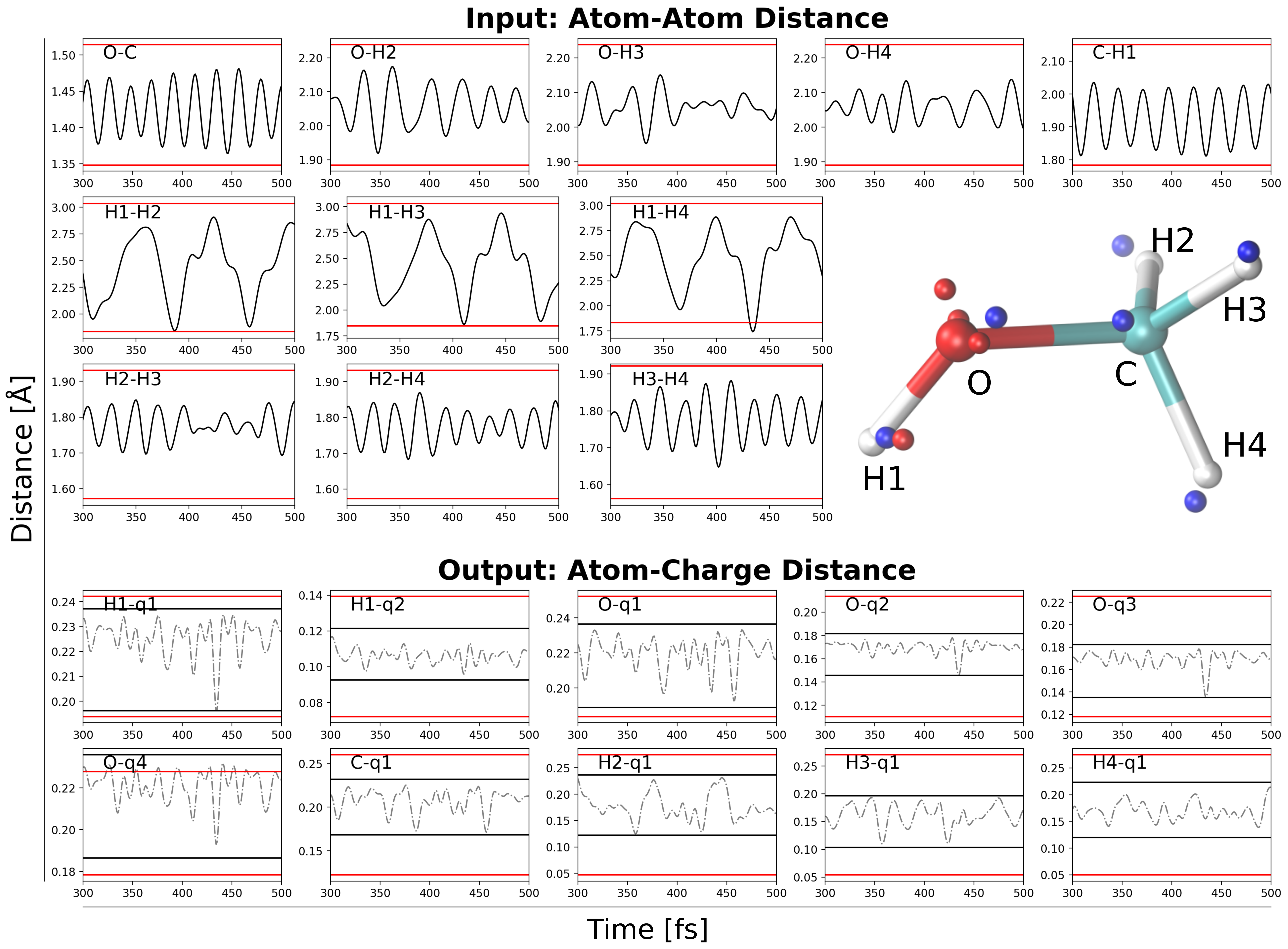}
    \caption{Input distance matrix values (top) for methanol,
      excluding SHAKEd bonds, and (bottom) output charge displacements
      versus saved simulation time step (0.001 ps between
      frames). Note that atom-charge distances are roughly one order
      of magnitude smaller than atom-atom distances.  Red horizontal
      lines indicate coverage of the training set; black horizontal
      lines indicate the bounds of values obtained during simulation.
      (inset) A ball-and-stick structure of methanol with atom labels
      and distributed charges are shown. Positive and negative charges
      are colored blue and red, respectively.}
    \label{fig:kmdcm-bounds}
\end{figure}

\noindent
It is also interesting to visualise the displacement dynamics of the
distributed charges themselves. Kernel-based interpolation methods
rely on establishing a convex hull of training examples to support
their predictions. For equilibrium MD simulations, the extent of the
conformational space needed to sample reliable structures may be
large, depending on the size of the molecule and the active degrees of
freedom. In the likely event that the molecule reaches a conformation
that is not present in the training set, the dynamics of the
distributed charges should be stable. Ensuring that the displacements
of the distributed charges do not move too far away from the atom
centers is important to prevent charges approaching too close to one
another, causing extreme Coulomb interactions, or overstabilizing
unphysical conformations.\\

The charge displacements during a gas phase methanol simulation are
compared to the displacements present in the training set, for
atom-atom distances, i.e. kernel input, which are mostly in the bounds
of the training examples (Figure \ref{fig:kmdcm-bounds}).  The top
panels show the intramolecular distances used as input to the kernel,
where the red horizontal lines report the maximum and minimum values
present in the training set. In the bottom panels, black horizontal
lines indicate the bounds of the charge displacements sampled during
the simulation.  All charges remained within the bounds of the
training distribution, with the exception of the fourth charge on the
oxygen atom (O-q4, Figure \ref{fig:kmdcm-bounds}) which moved slightly
($\sim 0.05$ \AA) outside of the bounds but is still smaller than the
$r_\mathrm{min}/2$ for the oxygen atom type of 1.79 \AA~in the CGenFF
force field. All input distances remained inside the bounds of the
training data, apart from the H1-H4 distance which reached a minimum
distance at $\sim 440$ ps which was unseen by the kernel. Sharp drops
in the displacements for some charges also occur at this time, as the
similarity to training examples becomes lower, causing the kernel
predictions to decay - although they remain within the bounds of the
distribution.\\

\section{Discussion and Conclusions}
The present work introduces a kernel-based kMDCM to describe
intramolecular charge redistribution and the conformational dependence
of the molecular ESP in a fashion that is suitable for molecular
(dynamics) simulations. As such, kMDCM is the full-dimensional
generalization to explicitly parametrized flexible MDCM (fMDCM) where
charge displacements $(u,v,w)$ depend, e.g., on a valence angle
$\theta$.\cite{MM.fmdcm:2022} In terms of performance, kMDCM improves
the accuracy by a factor of two compared with a conformationally
averaged PC-based model for methanol. To put this into perspective it
should be mentioned that typical empirical force fields with fixed
atom-centered PCs do not use conformationally averaged
models\cite{cgenff,opls:1996,gromos:2005,amber:2004} and improvements
relative to such models are expected to be even larger. Similarly, for
the molecular dipole moment a considerable conformational dependence
was observed which can be best captured by kMDCM, followed by MDCM and
PC models fit to ensembles of structures. However, it is notable that
for the dipole moment (see Figure \ref{fig:dipoles}A) the PC-based
model (red) overestimates the reference dipole moment from electronic
structure calculations by a constant amount with a small tilt away
from the diagonal towards higher values, whereas the two MDCM-based
models overestimate small dipole moments and underestimate the largest
values. It is also demonstrated here that meaningful and
energy-conserving condensed-phase simulations for 2000 water molecules
on the nanosecond time scale can be carried out. Thus, the present
results provide a strong basis for the validity of kernel-based
fluctuating charge models to systematically extend the scope of
empirical force fields.\\

\begin{figure}[h!]
    \centering
\includegraphics[width=0.96\textwidth]{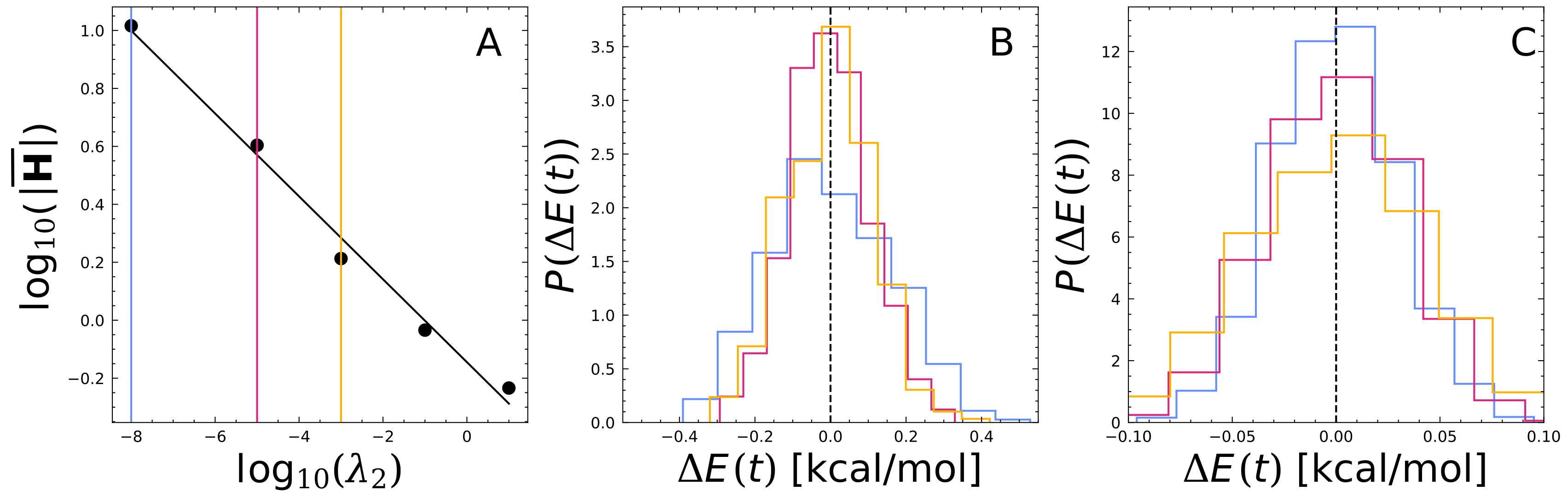}
    \caption{Panel A: The log-log relationship between the complexity
      parameter ($\lambda_2$) and the average norm of the Hessian
      (evaluated on off grid-points, for a single local charge
      position). kMDCM models fit using $\lambda_2$ of $10^{-8}$
      (blue), $10^{-5}$ (red), $10^{-3}$ (orange) provide similar test
      set performance. Panels B and C: The distribution in the total
      energy fluctuations using a time step of 0.2 fs and 0.1 fs is
      largely insensitive to the choice of $\lambda_2$.}
    \label{fig:lambda2}
\end{figure}

\noindent
As with all machine learning-based methods, there are system
parameters (hyperparameters) that are worthwhile to be analyzed in
more detail. This showed, for example, that the expressivity of the
model is largest for small values of $\lambda_2$ and increasing this
hyperparameter ties the model more to the average data. In other
words, $\lambda_2$ directly influences the smoothness of the
model. This can also be characterized by the average norm of the
Hessian\cite{Hug:2005,Stuetzle:2003} evaluated on off grid-points in
Figure \ref{fig:lambda2}A. Small values of $\lambda_2$ cause charges
to move more rapidly for conformations outside of the training set
whereas larger values of $\lambda_2$ lead to more slowly-varying
charge displacements without necessarily degrading the accuracy of the
model in representing the conformational dependence of the ESP.\\

It is interesting to consider whether such effects impact the quality
of MD simulations, e.g. with respect to conservation of the total
energy in $NVE$ simulations. For this, models for water with
$\lambda_2 = 10^{-8}, 10^{-5}, 10^{-3}$ were constructed. Three
different time steps, $\Delta t = 0.5, 0.2, 0.1$ fs, were used in the
MD simulations. The distributions of the energy fluctuations around
the mean are reported in Figures \ref{fig:lambda2}B and C for $\Delta
t = 0.2$ fs and 0.1 fs. The simulations were carried out for the water
box with 2000 monomers, used the RKHS representation for
intramolecular interactions and kMDCM for the electrostatics, with the
bonds involving hydrogen atoms free to move (flexible monomers). The
distributions $P(\Delta E)$ are largely insensitive to the value of
$\lambda_2$ whereas doubling the time step from 0.1 to 0.2 fs causes
the distribution to widen somewhat. For $\Delta t = 0.5$ fs the
simulations showed a monotonic increase in the total energy. Repeat
simulations under the same conditions using PCs instead of kMDCM and
the RKHS for the intramolecular contributions yielded comparable
distributions and an increase in the total energy for $\Delta t = 0.5$
fs which is due to the flexible OH-bonds. This was confirmed by
running simulations with constrained OH bond lengths with $\Delta t =
0.5$ and 1.0 fs using kMDCM which again conserved total energy. Hence,
kMDCM allows robust, energy-conserving MD simulations.\\

\noindent
The extrapolation capabilities of kMDCM were found to be satisfactory
and suggest that the method can also be applied to larger and more
flexible molecules, see Figure \ref{fig:si-ku}. Further improvement of
the performance of kMDCM can be expected by changing the kernel
functions from a Gaussian to other possible
kernels.\cite{MM.rkhs:2017} Another extension is afforded by using the
off-center PCs to describe external polarization. This will
amount to evaluating kMDCM for a given structure and then reposition
the off-center charges depending on the external electric field either
in a self-consistent or a non-self-consistent manner.\\

\noindent
In conclusion, kMDCM is a versatile model to describe intramolecular
charge redistribution depending on molecular conformations. The
formulation lends itself to be used in meaningful MD simulations and
the approach was implemented in and validated with the CHARMM
molecular simulation program.\\

\section*{Supporting Information Available}
The codes for this work are available at
\url{https://github.com/MMunibas/kMDCM}.

\begin{acknowledgement}
This work was supported by the Swiss National Science Foundation
through grants $200020\_219779$ and $200021\_215088$ and the
University of Basel, and by the European Union's Horizon 2020 research
and innovation program under the Marie Sk{\l}odowska-Curie grant
agreement No 801459-FP-RESOMUS.

\end{acknowledgement}

\bibliography{refs}

\end{document}